\documentclass[10pt, final, conference, letterpaper, onecolumn, twoside, romanappendices]{./IEEEtran}
\usepackage{cite}
\usepackage{graphicx}
\usepackage{subcaption}
\usepackage{amsmath}
\usepackage{amssymb}
\usepackage{mathtools}
\usepackage{bm}
\usepackage{epsfig}
\usepackage{times}
\usepackage{color}
\usepackage{url}
\usepackage{array}
\usepackage{multirow}
\usepackage{rotating}
\usepackage{extarrows}
\usepackage{mathabx} 
\usepackage{wasysym}

\newtheorem{pavikl}{\textbf{Lemma}}

\newtheorem{pavike}{\textbf{Example}}

\newcommand{\argmax}{\operatornamewithlimits{argmax}}

\title{Secure Analog Network Coding in Layered Networks}%
\author{\IEEEauthorblockN{Tulika Agrawal and Samar Agnihotri}%
\IEEEauthorblockA{School of Computing and Electrical Engineering, Indian Institute of Technology Mandi, HP - 175$\,$001, India}%
Email: tuliks.agrawal@gmail.com, samar.agnihotri@gmail.com%
}

\begin{document}

\maketitle

\begin{abstract}
We consider a class of Gaussian layered networks where a source communicates with a destination through $L$ intermediate relay layers with $N$ nodes in each layer in the presence of a single eavesdropper which can overhear the transmissions of the nodes in any one layer. The problem of maximum secrecy rate achievable with
analog network coding for a unicast communication over such layered wireless relay networks with directed links is considered. A relay node performing analog network coding scales and forwards the signals received at its input. The key contribution of this work is a lemma that provides the globally optimal set of scaling factors for the nodes that maximizes the end-to-end secrecy rate for a class of layered networks. We also show that in the high-SNR regime, ANC achieves secrecy rates within a constant gap of the cutset upper bound on the secrecy capacity. To the best of our knowledge, this work offers the first characterization of the performance of secure ANC in multi-layered networks in the presence of an eavesdropper.
\end{abstract}

\section{Introduction}
\label{sec:Intro}
Wireless communication, by its inherent broadcast nature, is vulnerable to eavesdropping by illegitimate receivers within communication range of the source. Wyner in \cite{075wyner}, for the first time, information-theoretically addressed the problem of secure communication in the presence of an eavesdropper and showed that secure communication is possible if the eavesdropper channel is a degraded version of the destination channel. The rate at which information can be transferred from the source to the intended destination while ensuring complete equivocation at the eavesdropper is termed as \textit{secrecy rate} and its maximum over all input probability distributions is defined as the \textit{secrecy capacity} of channel. Later, \cite{078cheongHellman} extended Wyner's result to Gaussian channels. These results are further extended to various models such as multi-antenna systems \cite{105paradaBlahut, 110khistiWornell}, multiuser scenarios \cite{108liuMaric, 108khisti_thesis}, fading channels \cite{108liangPoorShamai, 108gopalaLaiGamal}.

An interesting direction of work on secure communication in the presence of eavesdropper(s) is one in which the source communicates with the destination via relay nodes \cite{108laiGamal, 109dongHanPetropuluPoor, 110zhangGursoy, 113yangLiMaChing, 115sarmaAgnihotriKuri, 116sarmaAgnihotriKuri}. Such work has considered various scenarios such as different relaying schemes (\textit{amplify-and-forward} and \textit{decode-and-forward}), constraints on total or individual relay power consumption, one or more eavesdroppers. However, except for a few specific scenarios, such work does not provide tight characterization of secrecy capacity or even optimal secrecy rate achievable with the given relaying scheme. Further, all previous work only considered secure communication scenarios where the source communicates with the legitimate destination(s) in two hops, over so called \textit{diamond network} \cite{101schein}.

We consider a multihop unicast communication over \textit{layered} network of relays in the presence of a single eavesdropper. The relays nodes are arranged in layers where all relays in a particular layer can communicate only with the relays in the next layer. The relay nodes, operating under individual power constraints, amplify-and-forward the signals received at their input. In this scenario, multiple relay nodes in each layer can cooperate to enhance the end-to-end achievable rate. Also, the signals transmitted simultaneously by the relays add in the air, thus providing an opportunity for the relays in the second layer onward to perform Analog Network Coding (ANC) on the received \textit{noisy sum} of these signals, where each relay merely amplifies and forwards this noisy sum \cite{107kattiGollakottaKatabi, 110maricGoldsmithMedard}.

The eavesdropper can overhear the transmissions from the relay nodes of any of the layers depending on its location. The objective is to maximize the rate of secure transmission from the source to the destination by choosing the optimal set of scaling factors for the ANC-relays, irrespective of the relays that the eavesdropper listens to. However, so far, there exists no closed-form expression or polynomial time algorithm to exactly characterize the optimal AF secrecy rate even for general two-hop (diamond) relay networks, except for a few specific cases where eavesdropper's channel is a degraded or scaled version of destination channel  \cite{116sarmaAgnihotriKuri} and characterizing the optimal AF secrecy rate for general layered network is an even harder problem than general diamond network. Thus, to get some insights into the nature of the optimal solution for such networks, we consider symmetric layered networks, where all channel gains between the nodes in two adjacent layers are equal, thus the nomenclature of these networks as \textit{``Equal Channel Gains between Adjacent Layers (ECGAL)''} networks \cite{112agnihotriJaggiChen}. We provide closed-form solutions for the optimal secure AF rate for such networks. We envision that these results may help us gain insight into the nature of the optimal solution and develop techniques which may further help in construction of low-complexity optimal schemes for general relay networks.

The eavesdropper being a passive entity, a realistic eavesdropper scenario is the one where nothing about the eavesdropper's channel is known, neither its existence, nor its channel state information (CSI). However, the existing work on secrecy rate characterization assumes one of the following: (1) the transmitter has prefect knowledge of the eavesdropper channel states, (2) \textit{compound channel:} the transmitter knows that the eavesdropper channel can take values from a finite set \cite{109liangKramerPoorShamai, 109kobayashiLiangShamaiDebbah, 109ekremUlukus}, and (3) \textit{fading channel:} the transmitter only knows distribution of the eavesdropper channel \cite{108liangPoorShamai, 108gopalaLaiGamal}. In this paper, we assume that the CSI of the eavesdropper channel is known perfectly for the following two reasons. First, this provides an upper bound to the achievable secure ANC rate for the scenarios where we have imperfect knowledge of the eavesdropper channel. For example, the lower (upper) bound on the compound channel problem can be computed by solving the perfect CSI problem with the worst (best) channel gain from the corresponding finite set. Further, this also provides a benchmark to evaluate the performance of achievability schemes in such imperfect knowledge scenarios. Second, this assumption allows us to focus on the nature of the optimal solution and information flow, instead of on complexities arising out of imperfect channel models.  

The key contribution of this work is the computation of the globally optimal set of scaling factors for the nodes that maximizes the end-to-end secrecy rate for a class of layered networks. We also show that in the high-SNR regime, ANC achieves secrecy rates within an explicitly computed constant gap of the cutset upper bound on the secrecy capacity. To the best of our knowledge, this work offers the first characterization of the performance of secure ANC in multi-layered networks in the presence of an eavesdropper.

\textit{Organization:} In Section~\ref{sec:sysMdl} we introduce the system model and formulate the problem of maximum secure ANC rate achievable in the proposed system model. In section~\ref{sec:OptBeta} we compute the optimal vector of scaling factors of the nodes of an ECGAL network when eavesdropper snoops on the transmissions of the nodes in any one of the $L$ layers. Then, in Section~\ref{sec:highSNRanalysis} we analyze the high-SNR behavior of the achievable secure ANC rate and show that it lies within a constant gap from the corresponding cutset upper bound on the secrecy capacity and Section~\ref{sec:numSim} we numerically validate these results. Finally, Section~\ref{sec:cnclsn} concludes the paper.

\section{System Model}
\label{sec:sysMdl}
\begin{figure}[!t]
\centering
\includegraphics[width=3.5in]{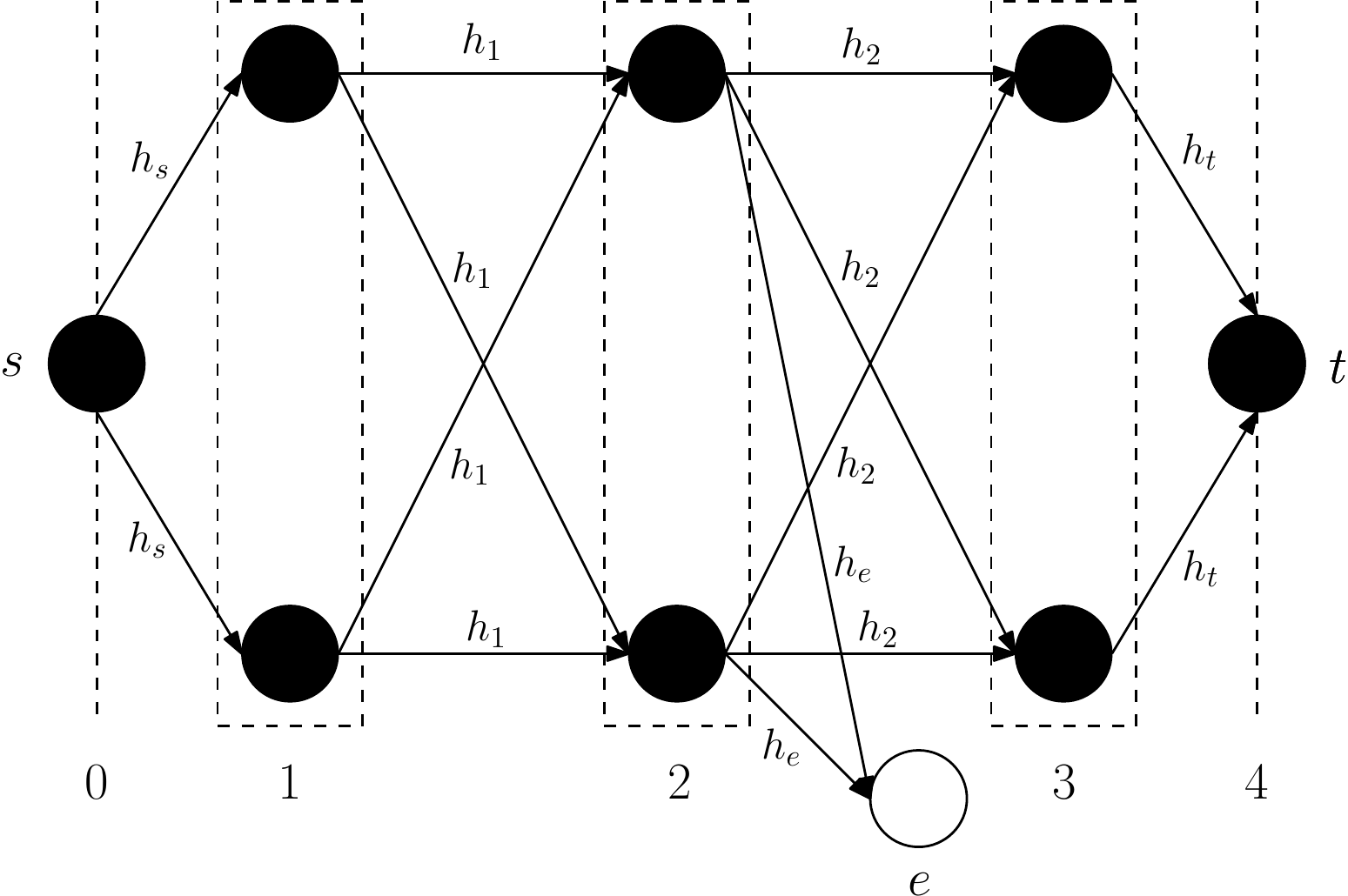}
\caption{An ECGAL network with 3 relay layers between the source $s$ and the destination $t$. Each layer contains two relay nodes. The eavesdropper overhears the transmissions from the relays in layer $2$.}
\label{fig:layrdNetExa}
\end{figure}

Consider a $(L+2)$-layer wireless network with directed links. The source $s$ is at layer `$0$', the destination $t$ is at layer `$L+1$' and the relays from the set $R$ are arranged in $L$ layers between them. The $l^{th}$ layer contains $n_l$ relay nodes, $\sum _{l-1}^{L} n_l = |R|$. The source $s$ transmits message signals to the destination $t$ via $L$ relay layers.  However, the signals transmitted by the relays in a layer are also overheard  by the eavesdropper $e$. An instance of such a network is given in Figure~\ref{fig:layrdNetExa}. Each node is assumed to have a single antenna and operate in full-duplex mode, \textit{e.g.} as in \cite{112agnihotriJaggiChen, 105gastparVetterli}.

At instant $n$, the channel output at node $i, i \in R \cup \{t, e\}$, is
\begin{equation}
\label{eqn:channelOut}
y_i[n] = \sum_{j \in {\mathcal N}(i)} h_{ji} x_j[n] + z_i[n], \quad - \infty < n < \infty,
\end{equation}
where $x_j[n]$ is the channel input of node $j$ in neighbor set ${\mathcal N}(i)$ of node $i$. In \eqref{eqn:channelOut}, $h_{ji}$ is a real number representing the channel gain along the link from the node $j$ to the node $i$ and constant over time (as in \cite{112agnihotriJaggiChen}, for example) and known (even for the eavesdropper channels) throughout the network \cite{109dongHanPetropuluPoor,110zhangGursoy}. All channel gains between the nodes in two adjacent layers are assumed to be equal, thus the nomenclature of these networks as \textit{``Equal Channel
Gains between Adjacent Layers (ECGAL)''} networks \cite{112agnihotriJaggiChen}. The source symbols $x_s[n], - \infty < n < \infty$, are i.i.d. Gaussian random variables with zero mean and variance $P_s$ that satisfy an average source power constraint, $x_s[n] \sim {\cal N}(0, P_s)$. Further, $\{z_i[n]\}$ is a sequence (in $n$) of i.i.d. Gaussian random variables with $z_i[n] \sim {\cal N}(0, \sigma^2)$. We assume that $z_i$ are independent of the input signal and of each other. We also assume that each relay's transmit power is constrained as:
\begin{equation}
\label{eqn:pwrConstraint}
E[x_i^2[n]] \le P, \quad i \in R, - \infty < n < \infty
\end{equation}

In ANC, each relay node amplifies and forwards the noisy signal sum received at its input. More precisely, relay node $i, i \in R$, at instant $n+1$ transmits the scaled version of $y_i[n]$, its input at time instant $n$, as follows
\begin{equation}
\label{eqn:AFdef}
x_i[n+1] = \beta_i y_i[n], \quad 0 \le \beta_i^2 \le \beta_{i,max}^2 = P/P_{R,i},
\end{equation}
where $P_{R,i}$ is the received power at node $i$ and choice of scaling factor $\beta_i$ satisfies the power constraint \eqref{eqn:pwrConstraint}.

Assuming equal delay along each path, for the network in Figure \ref{fig:layrdNetExa}, the copies of the source signal ($x_s[.]$) and noise signals ($z_i[.]$), respectively, arrive at the destination and the eavesdropper along multiple paths of the same delay. Therefore, the signals received at the destination and eavesdropper are free from intersymbol interference (ISI). Thus, we can omit the time indices and use equations \eqref{eqn:channelOut} and \eqref{eqn:AFdef} to write the input-output channel between the source $s$ and the destination $t$ as
\begin{equation}
\label{eqn:destchnl}
y_t = \left[\sum\limits_{(i_1,...,i_L) \in K_{st}}\!\!\!\!\!\!\!\!\! h_{s,i_1}\beta_{i_1}h_{i_1, i_2}...h_{i_{L-1}, i_L}\beta_{i_L} h_{i_L, t}\right] x_s + \sum\limits_{l=1}^L \sum\limits_{j-1}^{n_l}\left[\sum\limits_{(i_1,...,i_{L-l}) \in K_{lj,t}} \!\!\!\!\!\!\!\!\!\beta_{lj} h_{lj,i_1}...\beta_{i_{L-l}} h_{i_{L-l},t}\right] z_{lj} + z_t
\end{equation}
where $K_{st}$ is the set of $L$-tuples of node indices corresponding to all paths from the source $s$ to the destination $t$ with path delay $L$. Similarly, $K_{lj,t}$ is the set of $L - l$- tuples of node indices corresponding to all paths from the $j^{th}$ relay of the $l^{th}$ layer to the destination with path delay $L - l + 1$.

We introduce modified channel gains as follows. For all the paths between the source and the  destination:
\begin{equation}
h_{st} = \sum\limits_{(i_1,...,i_L) \in K_s} h_{s,i_1}\beta_{i_1}h_{i_1, i_2}...h_{i_{L-1}, i_L}\beta_{i_L} h_{i_L, t}
\end{equation}
For all the paths between the $j^{th}$ relay of the $l^{th}$ layer to the destination $t$ with path delay $L - l + 1$:
\begin{equation}
h_{lj,t} = \sum\limits_{(i_1,...,i_{L-l}) \in K_{lj}} \beta_{lj} h_{lj,i_1}...\beta_{i_{L-l}} h_{i_{L-l},t}
\end{equation}

In terms of these modified channel gains, the source-destination channel in \eqref{eqn:destchnl} can be written as: 
\begin{equation}
\label{eqn:destchnlmod}
y_t = h_{st} x_s + \sum_{l=1}^{L} \sum_{j=1}^{n_l} h_{lj,t} z_{lj} + z_t,
\end{equation}

Similarly, assuming that the eavesdropper is overhearing the transmissions of the relays in the layer $E, 1\le E \le L$, the input-output channel between the source and the eavesdropper can be written as 
\begin{equation}
\label{eqn:evechnlmod}
y_e = h_{se} x_s + \sum_{l=1}^{E} \sum_{j=1}^{n_l} h_{lj,e} z_{lj} + z_t,
\end{equation}

The secrecy rate at the destination for such a network model can be written as \cite{075wyner},  
$R_s(P_s)= [I(x_s;y_t)-I(x_s;y_e)]^+$,
where $I(x_s;y)$ represents the mutual information between random variable $x_s$ and $y$ and $[u]^+=\max\{u,0\}$.

The secrecy capacity is attained for the Gaussian channels with the Gaussian input $x_s \sim \mathcal{N}(0,P_s)$, where $\mathbf{E}[x_s^2] = P_s$, \cite{078cheongHellman}. Therefore, for a given network-wide scaling vector $\bm{\beta} = (\beta_{li})_{1 \le l \le L, 1 \le i \le n_l}$, the optimal secure ANC rate for the channels in \eqref{eqn:destchnlmod} and \eqref{eqn:evechnlmod} can be written as the following optimization problem.
\begin{subequations}
\label{eq:optSecrate}
\begin{align}
R_s(P_s) &=  \max_{\bm{\beta}} \left[R_t(P_s,\bm{\beta}) - R_e(P_s,\boldsymbol{\beta})\right]\\
         &=  \max_{\boldsymbol{\beta}} \left[\frac{1}{2}\log\frac{1+SNR_t(P_s,\bm{\beta})}{1+SNR_e(P_s,\bm{\beta})}\right],
\end{align}  
\end{subequations}
where $SNR_t(P_s,\bm{\beta})$, the signal-to-noise ratio at the destination $t$ is:
\begin{equation}
\label{eqn:snrt}
SNR_t(P_s,\bm{\beta}) = \frac{P_s}{\sigma^2}\frac{h_{st}^2}{1 + \sum_{l=1}^{L} \sum_{j=1}^{n_l} h_{lj,t}^2}
\end{equation}
and similarly, $SNR_e(P_s,\bm{\beta})$ is
\begin{equation}
\label{eqn:snre}
SNR_e(P_s,\bm{\beta}) = \frac{P_s}{\sigma^2}\frac{h_{se}^2}{1 + \sum_{l=1}^{E} \sum_{j=1}^{n_l} h_{lj,e}^2}
\end{equation}

Given the monotonicity of the $\log(\cdot)$ function, we have
\begin{align}
\bm{\beta}_{opt} &= \argmax_{\bm{\beta}} \left[R_t(P_s,\bm{\beta}) - R_e(P_s,\boldsymbol{\beta})\right] \nonumber\\
                 &= \argmax_{\bm{\beta}} \frac{1+SNR_t(P_s,\bm{\beta})}{1+SNR_e(P_s,\bm{\beta})} \label{eqn:eqProb}
\end{align}

\section{The Optimal Secure ANC Rate Analysis}
\label{sec:OptBeta}
In this section, we analyze the optimal secure ANC rate problem in \eqref{eq:optSecrate} or \eqref{eqn:eqProb} first for diamond networks and then for ECGAL  networks.

\begin{figure}[!t]
\centering
\includegraphics[width=3.0in]{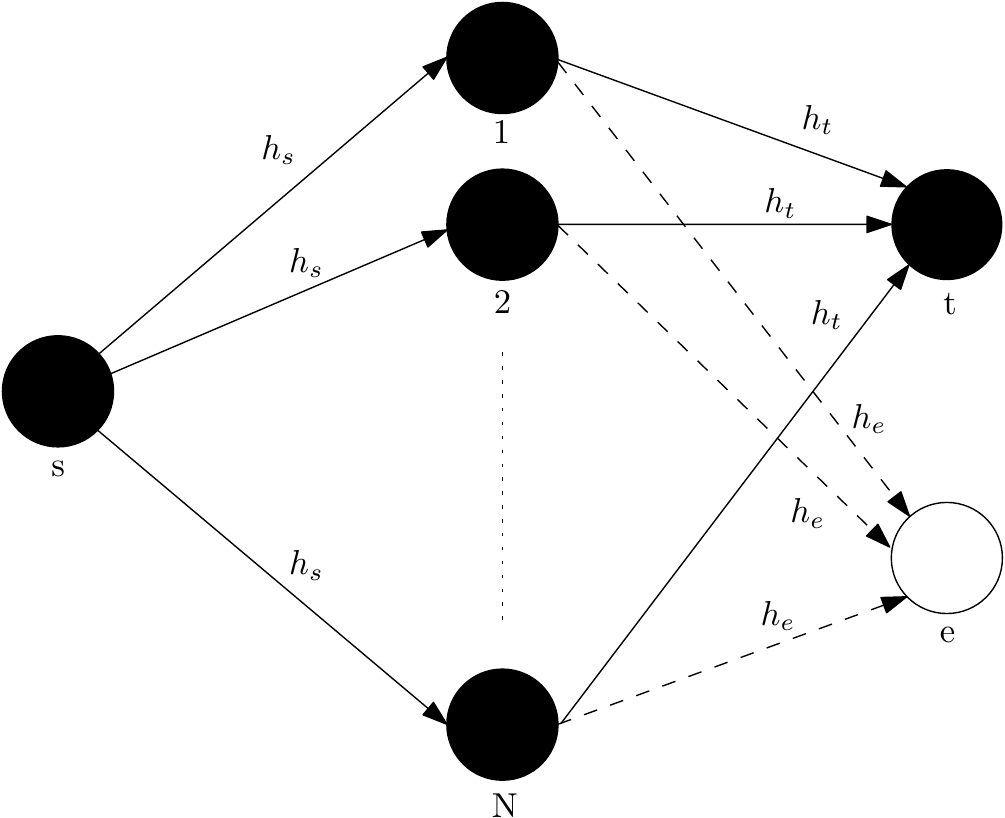}
\caption{A symmetric $N$ relay diamond network with an eavesdropper.}
\label{fig:diamondNet}
\end{figure}

\subsection{Symmetric Diamond Networks}
\label{subsec:diamondNetOptBeta}
Consider a symmetric diamond with $N$ relay nodes arranged in a layer between the source and the destination as shown in Figure~\ref{fig:diamondNet}. Using \eqref{eqn:snrt} and \eqref{eqn:snre}, the $SNR_t$ and $SNR_e$ in this case are:
\begin{align*}
SNR_t = \frac{P_s h_s^2 }{\sigma^2} \frac{(\sum_{i=1}^{N} \beta_i )^2 h_t^2}{1 + \left(\sum_{i=1}^{N} \beta_i^2 \right) h_t^2 } &&\mbox{ and }&& SNR_e = \frac{P_s h_s^2 }{\sigma^2} \frac{(\sum_{i=1}^{N} \beta_i )^2 h_e^2}{1 + \left(\sum_{i=1}^{N} \beta_i^2 \right) h_e^2 }
\end{align*}
\begin{pavikl}
\label{lemma:diamondNetReducedBeta1}
For symmetric diamond network, $\bm{\beta}_{opt}$ in \eqref{eqn:eqProb} is:
\begin{align*}
\beta_{1,opt},\cdots,\beta_{N,opt}=\beta_{opt}=
			\begin{cases}
               \min(\beta_{max}^2, \beta_{glb}^2), \quad \mbox{if } h_{t}>h_e,\\
               0, \quad \mbox{otherwise}
            \end{cases}
\end{align*}
where
\begin{align*}
\beta_{glb}^2 &= \sqrt{\frac{1}{N^2 h_t^2 h_e^2 \left(1+N\frac{P_s h_s^2}{\sigma^2} \right)}}
\end{align*}
\end{pavikl}
\begin{IEEEproof}
Please refer to Appendix~\ref{appndx:lemma0}.
\end{IEEEproof}
Here, it is assumed that the eavesdropper chooses to snoop on all the nodes of the layer which is an optimal strategy in the symmetric layered networks for the eavesdropper as we prove later. However, in general this may not be the case. The eavesdropper can choose to snoop on fewer nodes and still get higher rate compared to the case when it snoops on all the nodes of a layer as illustrated in the following example.
\begin{pavike}
Consider the relay network shown in Figure~\ref{fig:counterex}. Let $h_s=0.6,\  h_t=0.3,\  h_{1e}=0.2,\  h_{2e}=0.6,\  h_{3e}=0.4$. Let $P_s=P_1=P_2=P_3=5$ and noise variance $\sigma^2=1.0$ at each node.

\begin{figure}[!t]
\centering
\includegraphics[width=2.5in]{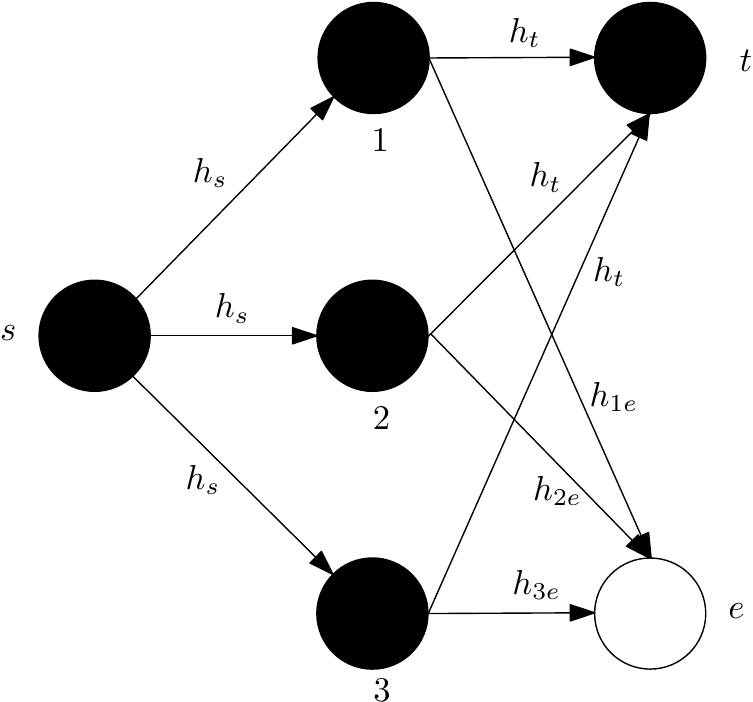}
\caption{3 relay diamond network with eavesdropper overhearing the transmissions of all the relay nodes.}
\label{fig:counterex}
\end{figure}
\textit{Case 1:} Eavesdropper chooses to snoop on all the relay nodes.\\
In this case we have the following secrecy rate maximization problem:
\begin{align*}
R_s &= \max_{\beta_1, \beta_2, \beta_3} \left\{ \frac{1}{2} \log\left(\!1\!\!+\!\!\frac{P_s h_s^2}{\sigma^2} \frac{\left(\beta_1\!+\!\beta_2\!+\!\beta_3\right)^2 h_t^2}{1\!+\!\left(\beta_1^2\!+\!\beta_2^2\!+\!\beta_3^2\right)h_t^2}\!\right) -  \frac{1}{2} \log\left(1\!\!+\!\!\frac{P_s h_s^2}{\sigma^2} \frac{\left(\beta_1 h_{1e} \!+\! \beta_2 h_{2e} \!+\! \beta_3 h_{3e}\right)^2}{1\!+\!\beta_1^2 h_{1e}^2 \!+\! \beta_2^2 h_{2e}^2 \!+\! \beta_3^2 h_{3e}^2}\right)\right\}\\
	&= \max_{\beta_1, \beta_2, \beta_3} \left\{ \frac{1}{2} \log\left(1\!+\! \frac{0.162\left(\beta_1\!+\!\beta_2\!+\!\beta_3\right)^2 }{1\!+\!\left(\beta_1^2\!+\!\beta_2^2\!+\!\beta_3^2\right)0.09}\right) -  \frac{1}{2} \log\left(1\!+\! \frac{1.8\left(0.2 \beta_1 \!+\! 0.6 \beta_2  \!+\! 0.4 \beta_3\right)^2}{1\!+\!0.04 \beta_1^2  \!+\! 0.36 \beta_2^2 \!+\! 0.16 \beta_3^2 }\right)\right\}
\end{align*}
The optimal solution of this problem is $\beta_1=\beta_{1,max} = 1.3363, \ \beta_2 = 0.0, \ \beta_3 = 0.0$. For these optimum values of $\beta$'s, the rate achievable at the eavesdropper is:
\begin{align*}
R_e &= \frac{1}{2} \log\left(1+ \frac{1.8\left(0.2 \beta_1 + 0.6 \beta_2  + 0.4 \beta_3\right)^2}{1+0.04 \beta_1^2  + 0.36 \beta_2^2 + 0.16 \beta_3^2 }\right) = 0.081749 \mbox{ bits/s/Hz}
\end{align*}
\textit{Case 2:} Eavesdropper chooses to snoop on relay nodes $2$ and $3$. \\
In this case we have the following secrecy rate maximization problem:
\begin{align*}
R_s &= \max_{\beta_1, \beta_2, \beta_3} \left\{ \frac{1}{2} \log\left(1+\frac{P_s h_s^2}{\sigma^2} \frac{\left(\beta_1+\beta_2+\beta_3\right)^2 h_t^2}{1+\left(\beta_1^2+\beta_2^2+\beta_3^2\right)h_t^2}\right) -  \frac{1}{2} \log\left(1+\frac{P_s h_s^2}{\sigma^2} \frac{\left(\beta_2 h_{2e} + \beta_3 h_{3e}\right)^2}{1+\beta_2^2 h_{2e}^2 +  \beta_3^2 h_{3e}^2}\right)\right\}\\
	&= \max_{\beta_1, \beta_2, \beta_3} \left\{ \frac{1}{2} \log\left(1+ \frac{0.162\left(\beta_1+\beta_2+\beta_3\right)^2 }{1+\left(\beta_1^2+\beta_2^2+\beta_3^2\right)0.09}\right)-  \frac{1}{2} \log\left(1+ \frac{1.8\left( 0.6 \beta_2  + 0.4 \beta_3\right)^2}{1+ 0.36 \beta_2^2 + 0.16 \beta_3^2 }\right)\right\}
\end{align*}
The optimal solution of this problem is $\beta_1=\beta_{1,max} = 1.3363, \ \beta_2 = 0.0, \ \beta_3 = 0.7298$. For these optimum values of $\beta$'s, the rate achievable at the eavesdropper is:
\begin{align*}
R_e &= \frac{1}{2} \log\left(1+ \frac{1.8\left( 0.6 \beta_2  + 0.4 \beta_3\right)^2}{1+ 0.36 \beta_2^2 + 0.16 \beta_3^2 }\right) = 0.095368 \mbox{ bits/s/Hz}
\end{align*}
From above it is clear that in the asymmetric diamond networks, the eavesdropper achieves a higher rate when it chooses to snoop on two nodes ($2$ and $3$) compared to the case where eavesdropper snoops on all the three nodes.\hfill\IEEEQEDclosed
\end{pavike}

Although, for general layered networks it is very difficult to find which subset of relay nodes the eavesdropper chooses to snoop on so as to maximize its rate; for symmetric networks it can be easily verified that the rate at the eavesdropper is maximized when it snoops on all the nodes of a layer. For instance, consider the scenario where all the nodes transmitting at maximum power is optimum with respect to secrecy rate maximization. The SNR at the eavesdropper when it chooses to snoop on $k$ nodes is given as
\begin{align}
SNR_e^k & = \frac{P_s h_s^2}{\sigma^2} \frac{k^2 \beta_{max}^2 h_e^2}{1+ k \beta_{max}^2 h_e^2}
\end{align}

Similarly, SNR at the eavesdropper when it chooses to snoop on $k+1$ nodes is
\begin{align}
SNR_e^{k+1} & = \frac{P_s h_s^2}{\sigma^2} \frac{(k+1)^2 \beta_{max}^2 h_e^2}{1+ (k+1) \beta_{max}^2 h_e^2}
\end{align}

Clearly,
\begin{align*}
SNR_e^{k+1}-SNR_e^k & = \frac{P_s h_s^2}{\sigma^2} \frac{{\beta}^{2} h_e^2 \left( {\beta}^{2} h_e^2 {k}^{2}+{\beta}^{2} h_e^2 k+2 k+1\right)}{\left( {\beta}^{2} h_e^2 k+1\right)  \left( {\beta}^{2} h_e^2 k+{\beta}^{2} h_e^2+1\right) }  \geq 0,
\end{align*}
i.e., $SNR_e^{k+1} \geq SNR_e^k$. Thus, for such scenarios eavesdropper achieves a higher rate when it chooses to snoop on more number of nodes and eventually it will snoop on maximum possible number of nodes to maximize its rate.

\begin{figure}[!t]
\centering
\includegraphics[width=4.5in]{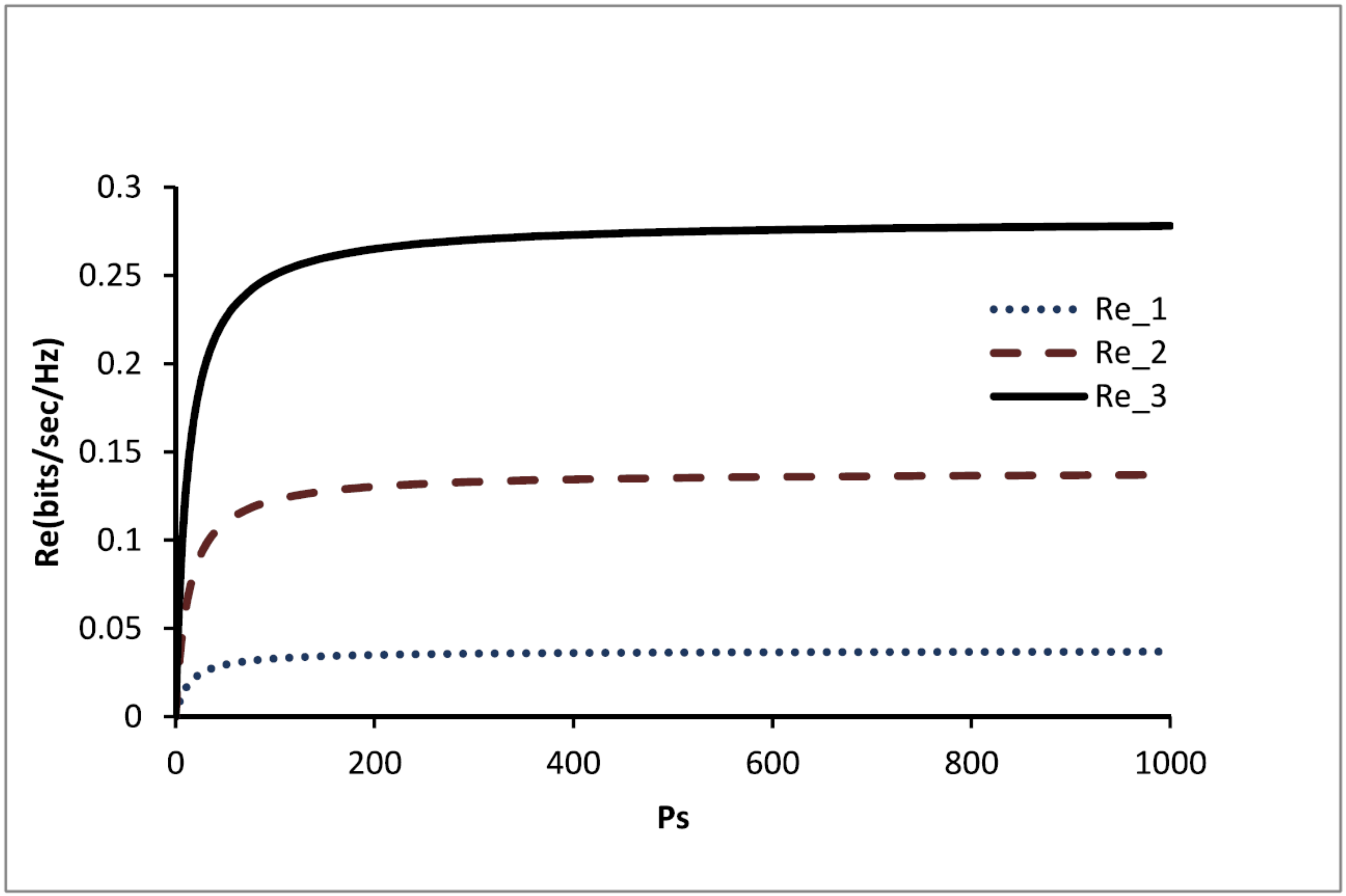}
\caption{Achievable rate at the eavesdropper when it snoops on different number of nodes of a symmetric diamond network of three relay nodes with $P=10.0,\ \sigma^2=1.0, \ h_s=0.278, \ h_t=0.379, \ h_e=0.073$.}
\label{fig:3relayDiamondNetRePlot}
\end{figure}

Figure~\ref{fig:3relayDiamondNetRePlot} shows the rates achievable at the eavesdropper $(R_{e,i})$ when it chooses to snoop on $i,\ i  \in \{1,2,3\}$ number of nodes of a 3 relay symmetric diamond network of the specified parameters with all the nodes transmitting at their corresponding optimum values so as to maximize the secrecy rate in each case. Here, again it can be seen that rate at the eavesdropper is increases when it snoops on all the nodes.
\subsection{ECGAL Layered Networks}
\label{subsec:lyrNetOptBeta}
In this subsection, we consider the optimal secure ANC rate problem in \eqref{eq:optSecrate} for ECGAL networks where the source communicates with the destination via $L$ intermediate relay layers with all channel gains between two adjacent layers being equal. For the sake of ease of representation, let there be $N$ relays in each layer. The eavesdropper overhears the transmission from the nodes in relay layer $M, 1 \le M \le L$. An instance of such a network is given in Figure~\ref{fig:layrdNetExa}.

Using \eqref{eqn:snrt} and \eqref{eqn:snre}, the $SNR_t$ and $SNR_e$ in this case are:
\begin{align*}
SNR_t &= \frac{P_s}{\sigma^2} \frac{h_s^2 H_{1,M-1}^2 \left(\sum_{n=1}^N \beta_{M,n}\right)^2 h_M^2 H_{M+1,L}^2}{\left[\left(\sum_{i=1}^{M-1} G_{i,M-1}^2\right)\left(\sum_{n=1}^N \beta_{M,n}\right)^2 h_M^2 +\left(\sum_{n=1}^N \beta_{M,n}^2\right)h_M^2\right] H_{M+1,L}^2 +\sum_{i=M+1}^L G_{i,L}^2+1}\\
SNR_e &= \frac{P_s}{\sigma^2} \frac{h_s^2 H_{1,M-1}^2 \left(\sum_{n=1}^N \beta_{M,n}\right)^2 h_e^2}{\left(\sum_{i=1}^{M-1} G_{i,M-1}^2\right)\left(\sum_{n=1}^N \beta_{M,n}\right)^2 h_e^2+\left(\sum_{n=1}^N \beta_{M,n}^2\right)\,h_e^2+1}
\end{align*}
where
\begin{align*}
H_{i,j}^2 &= \prod_{k=i}^j \left(\sum_{n=1}^N \beta_{k,n}\right)^2 h_k^2\\
G_{i,j}^2 &= \left(\sum_{n=1}^N \beta_{i,n}^2\right) h_i^2 \prod_{k=i+1}^{j} \left(\sum_{n=1}^N \beta_{k,n}\right)^2 h_k^2
\end{align*}

\begin{pavikl}
\label{lemma:lyrdNetReducedBeta1}
For the ECGAL layered networks, $\bm{\beta}_{opt}$ in \eqref{eqn:eqProb} is:
\begin{equation*}
\bm{\beta}_{opt} = (\bm{\beta}_{1,opt}, \ldots, \bm{\beta}_{M,opt}, \bm{\beta}_{M+1,max}, \ldots, \bm{\beta}_{L,max})
\end{equation*}
\end{pavikl}
\begin{IEEEproof}
Please refer to Appendix~\ref{appndx:lemma4}.
\end{IEEEproof}

Introduce the following parameters
\begin{align*}
E &= h_s^2 H_{1,M-1}^2 \\
F &= \sum_{i=1}^{M-1} G_{i,M-1}^2 \\
\alpha &= \prod_{i=M+1}^{L}\left(\sum_{n=1}^N\sqrt{P_{i,n}}\right)^2 h_i^2 \\
\lambda &= \sum_{i=M+1}^L \lambda_{i} \left(\sum_{n=1}^N P_{i+1,n}\right) \prod_{j=i+2}^{L}\left(\sum_{n=1}^N\sqrt{P_{i,n}}\right)^2 h_j^2 \\
\mu &= \prod_{i=M+1}^{L}\left(\sum_{n=1}^N\sqrt{P_{i,n}}\right)^2 h_i^2 + \sum_{i=M+1}^L \mu_{i} \left(\sum_{n=1}^N P_{i+1,n}\right) \prod_{j=i+2}^{L}\left(\sum_{n=1}^N\sqrt{P_{i,n}}\right)^2 h_j^2\\
    &= \alpha + \sum_{i=M+1}^L \mu_{i} \left(\sum_{n=1}^N P_{i+1,n}\right) \prod_{j=i+2}^{L}\left(\sum_{n=1}^N\sqrt{P_{i,n}}\right)^2 h_j^2 \\
\nu &= \sum_{i=M}^L \nu_{i} \left(\sum_{n=1}^N P_{i+1,n}\right) \prod_{j=i+2}^{L}\left(\sum_{n=1}^N\sqrt{P_{i,n}}\right)^2 h_j^2, \quad \nu_M = 1,
\end{align*}
where
\begin{align*}
\lambda_{M+1} &= P_s (s_{M+1} + \sigma^2 n_{M+1}), \quad s_{M+1} = 1, n_{M+1} = 0 &\\
\lambda_{i} &= P_s\left[s_{i-1} \left(\left(\sum_{n=1}^N\!\!\sqrt{P_{i-1,n}}\right)^2\!\! h_{i-1}^2\!\!+\!\sigma^2\right)\!\! +\! \sigma^2 n_{i-1} \left(\left(\sum_{n=1}^N P_{i-1,n}\right) h_{i-1}^2 \!\!+\!\sigma^2 \right)\right], & i \in \{M+2, \ldots, L\}\\
\mu_{M+1} &= \sigma^2 (s_{M+1} + \sigma^2 n_{M+1}), \quad s_{M+1} = 1, n_{M+1} = 0 &\\
\mu_{i} &= \sigma^2 \left[s_{i-1} \left(\left(\sum_{n=1}^N\!\!\sqrt{P_{i-1,n}}\right)^2\!\! h_{i-1}^2\!\!+\!\sigma^2\right)\!\! +\! \sigma^2 n_{i-1} \left(\left(\sum_{n=1}^N P_{i-1,n}\right) h_{i-1}^2 \!\!+\!\sigma^2 \right)\right], & i \in \{M+2, \ldots, L\}\\
\nu_{M+1} &= (s_{M+1} + \sigma^2 n_{M+1}), \quad s_{M+1} = 0, n_{M+1} = 1&\\
\nu_{i} &= \sigma^2 \left[s_{i-1} \left(\left(\sum_{n=1}^N\!\!\sqrt{P_{i-1,n}}\right)^2\!\! h_{i-1}^2\!\!+\!\sigma^2\right)\!\! +\! \sigma^2 n_{i-1} \left(\left(\sum_{n=1}^N P_{i-1,n}\right) h_{i-1}^2 \!\!+\!\sigma^2 \right)\right], & i \in \{M+2, \ldots, L\}
\end{align*}

Using the preceding lemma and the above parameters, the problem
\begin{equation*}
\bm{\beta}_{opt} = \argmax_{\bm{\beta}} \frac{1+SNR_t}{1+SNR_e}
\end{equation*}
is reduced to the following subproblem
\begin{equation}
\label{eqn:secRate4lemma2}
(\bm{\beta}_{1,opt}, \ldots, \bm{\beta}_{M,opt}) = \argmax_{(\bm{\beta}_{1}, \ldots, \bm{\beta}_{M})} \frac{1+SNR_t|_{\bm{\beta}_{M+1:L,max}}}{1+SNR_e},
\end{equation}
where for a given network-wide vector of scaling factors $(\bm{\beta}_1, \ldots, \bm{\beta}_M, \bm{\beta}_{M+1,max}, \ldots, \bm{\beta}_{L,max})$, the received SNRs at the destination and the eavesdropper are, respectively
\begin{align*}
SNR_t|_{\bm{\beta}_{M+1:L,max}} &= \frac{P_s}{\sigma^2} \frac{A \left(\sum_{n=1}^N \beta_{M,n}\right)^2  h_M^2}{B \left(\sum_{n=1}^N \beta_{M,n}\right)^2 h_M^2 + C\left(\sum_{n=1}^N \beta_{M,n}^2\right) h_M^2 + D}\\
SNR_e &= \frac{P_s}{\sigma^2} \frac{E \left(\sum_{n=1}^N \beta_{M,n}\right)^2 h_e^2}{F \left(\sum_{n=1}^N \beta_{M,n}\right)^2 h_M^2 + \left(\sum_{n=1}^N \beta_{M,n}^2\right) h_M^2 + 1}
\end{align*}
with
\begin{align*}
A &= \alpha E\\
B &= \lambda E + \mu F\\
C &= \mu\\
D &= \nu
\end{align*}

\begin{pavikl}
\label{lemma:lyrdNetReducedBeta2}
For ECGAL layered networks, the subvector $(\beta_{M,1,opt}, \ldots, \beta_{M,N,opt})$ of the optimum scaling vectors for the nodes in the $M^{th}$ layers for given sub-vector $(\bm{\beta}_1, \ldots, \bm{\beta}_{M-1})$ is:
\begin{equation*}
\beta_{M,1,opt}^2 = \ldots = \beta_{M,N,opt}^2 = \beta_{M,opt}^2 = 
                 \begin{cases}
                   \min(\beta_{M,max}^2, \beta_{M,glb}^2), \quad \mbox{if } (h_M^{2}\,\alpha-h_e^{2}\,\nu) > 0,\\
                   0, \quad \mbox{otherwise}
                 \end{cases}
\end{equation*}
where
\begin{equation*}
\beta_{M,glb}^2 = \frac{|{\mathcal B}|}{2\,|{\mathcal A}|}\left(\sqrt{1 + \frac{4\,|{\mathcal A}|\,{\mathcal C}}{{\mathcal B}^2}} - 1\right)
\end{equation*}
with
\begin{align*}
{\mathcal A} & = 4\,h_M^{2}\,h_e^{2}\bigg\{h_e^{2}\,\alpha\,\nu\,(2\,F+1)\,\left(2\,F+1+2\,\frac{P_s}{\sigma^2}\,E\right)\\
				&\quad -h_M^{2}\,[2\,\lambda E +(2\,F+1)\mu]\,\left[(2\,F+1)\mu+2\,\left(\lambda +\alpha\,\frac{P_s}{\sigma^2}\right) E\,\right]\bigg\}\\
{\mathcal B} & = 4\,h_M^{2}\,h_e^{2}\,\nu\,[(\alpha-\mu)\,(2\,F+1)-2\,\lambda E]\\
{\mathcal C} & = \nu\,(h_M^{2}\,\alpha-h_e^{2}\,\nu)
\end{align*}
\end{pavikl}
\begin{IEEEproof}
Please refer to Appendix~\ref{appndx:lemma5}.
\end{IEEEproof}

\begin{pavikl}
\label{lemma:lyrdNetReducedBeta3}
For ECGAL layered networks,
\begin{equation*}
(\bm{\beta_{1,opt}}, \ldots, \bm{\beta_{M-1,opt}}) = (\bm{\beta_{1,max}}, \ldots, \bm{\beta_{M-1,max}}),
\end{equation*}
where
\begin{align*}
\beta_{1,n,max}^2 &= \frac{P_{1,n}}{P_{s} h_s^2 + \sigma^2}, n \in \{1, \ldots, N\}\\
\beta_{i,n,max}^2 &= \frac{P_{i,n}}{P_{R_x,i}}, \quad i \in \{2, \ldots, M-1\},  n \in \{1, \ldots, N\}
\end{align*}
with
\begin{align*}
P_{Rx,i} &= P_s h_s^2 H_{1,i-1}^2 +\left[\sum_{j=1}^{i-1} G_{j,i-1}^2+1\right] \sigma^2
\end{align*}
\end{pavikl}
\begin{IEEEproof}
Please refer to Appendix~\ref{appndx:lemma6}.
\end{IEEEproof}

In short, Lemma~\ref{lemma:lyrdNetReducedBeta1}-\ref{lemma:lyrdNetReducedBeta3} together establish that for the \textit{ECGAL} layered networks with $L$ relays, the optimal vector of the scaling factors that maximizes the secure ANC rate is $\bm{\beta}_{opt} = (\bm{\beta}_{1,max}, \ldots, \bm{\beta}_{M,opt}, \bm{\beta}_{M+1,max}, \ldots, \bm{\beta}_{L,max})$, where $\beta_{M,opt}$ is given by Lemma~\ref{lemma:lyrdNetReducedBeta2}.

\section{High-SNR Analysis of Achievable ANC Secrecy Rate in ECGAL Networks}
\label{sec:highSNRanalysis}
We define a wireless layered network to be in high SNR regime if
\begin{equation*}
\min_{i \in \{1, \ldots, L\}} SNR_i \ge \frac{1}{\delta}
\end{equation*}
for some small $\delta \ge 0$. Here, $SNR_i$ is the signal-to-noise ratio at the input of any of the relay nodes in the $i^{th}$ layer.

Assume that each relay node in layer $i$ uses the amplification factor
\begin{equation*}
\beta_i^2 = \frac{P}{(1+\delta) P_{R_i,max}}, i \in \{1, \ldots, L\},
\end{equation*}
where $P_{R_i,max}$ is the maximum  received signal power at any  relay node in the $i^{th}$ layer which in this case is equal to $N^2 P h_{i-1}^2$ as each relay node receives the transmissions from $N$ relay nodes in the previous layer with maximum transmit power constrained by $P$. It should be noted that $\beta_i$ is such that the maximum power constraint \eqref{eqn:pwrConstraint} is satisfied at each node as 
\begin{align*}
\beta_{i,max}^2 = \frac{P}{P_{R_i}+P_{z_i}+\sigma^2} = \frac{P}{\left(1+\frac{1}{SNR_i}\right)P_{R_i}} \geq \frac{P}{(1+\delta)P_{R_i,max}} \quad [\mbox{Since }1/SNR_i \leq \delta, P_{R_i} \leq P_{R_i,max}]
\end{align*}
Here, $P_{R_i}$ and $P_{z_i}$ are the received signal and noise powers at the input of the node $i$, respectively. Note that as $\delta \rightarrow 0, \ \beta_i^2 \rightarrow \beta_{i,max}^2$.

We now analyze the secrecy rate achievable with these scaling factors. However, before discussing the achievable secrecy rate, we discuss an upper bound on the secrecy capacity of such a network. For the source-destination(eavesdropper) path, an upper bound on the capacity is given by the capacity of the Gaussian multiple-access channel between the relays in the $L^{th}$ layer and the destination (the eavesdropper). This results in the following upper bound on the secrecy capacity of such
networks:
\begin{equation*}
C_{cut} = \frac{1}{2}\log\left[\frac{1 + P_t/\sigma^2}{1 + P_e/\sigma^2}\right],
\end{equation*}
where $P_t = N^2 P_L h_t^2$ and $P_e = N^2 P_L h_e^2$.

The power of the source signal reaching the destination $t$ is:
\begin{align}
P_{s,t} &= P_s h_s^2 \left(\prod_{i=1}^{L-1} N^2\beta_i^2 h_i^2\right) N^2 \beta_L^2 h_t^2 = \frac{N^2 P_L h_t^2}{(1+\delta)^L} \label{eqn:source_power_at_dstntn}
\end{align}

The total power of noise reaching the destination $t$ from all relay nodes:
\begin{equation}
P_{z,t} = \sum_{i=1}^L P_{z,t}^i = \sigma^2\left(\sum\limits_{i=1}^{L-1}N \beta_i^2 h_i^2 \left(\prod\limits_{j=i+1}^{L-1}(N \beta_j h_j)^2\right)N^2 \beta_L^2 h_t^2 + N \beta_L^2 h_t^2\right)
\label{eqn:noise_power_at_dstntn}
\end{equation}
where $P_{z,t}^i$ is the noise power reaching the destination form nodes in $i^{th}$ layer.

Now,
\begin{align*}
&&P_{z,t}^1 &= \sigma^2 N \beta_1^2 h_1^2 \left(\prod\limits_{j=2}^{L-1}N^2\beta_j^2 h_j^2\right) N^2 \beta_L^2 h_t^2 = \frac{\sigma^2}{P_s h_s^2} \frac{N P h_t^2}{(1+\delta)^L} \leq \frac{\delta}{(1+\delta)^L}N P h_t^2 &&\\
&&P_{z,t}^2 &= \sigma^2 N \beta_2^2 h_1^2 \left(\prod\limits_{j=3}^{L-1}N^2\beta_j^2 h_j^2\right) N^2 \beta_L^2 h_t^2 = \frac{\sigma^2}{N^2 P h_1^2} \frac{N P h_t^2}{(1+\delta)^{L-1}} \leq \frac{\delta}{(1+\delta)^{L-1}}N P h_t^2 &&\\
&& \vdots & &&\\
&&P_{z,t}^i &= \sigma^2 N \beta_i^2 h_i^2 \left(\prod\limits_{j=i+1}^{L-1}N^2\beta_j^2 h_j^2\right) N^2 \beta_L^2 h_t^2 = \frac{\sigma^2}{N^2 P h_{i-1}^2} \frac{N P h_t^2}{(1+\delta)^{L-i+1}} \leq \frac{\delta}{(1+\delta)^{L-1}}N P h_t^2 &&\\
&& \vdots & &&\\
&&P_{z,t}^L &=  \sigma^2 N \beta_L^2 h_t^2 = \frac{\sigma^2}{N^2 P h_{L-1}^2} \frac{N P h_t^2}{(1+\delta)} \leq \frac{\delta}{(1+\delta)}N P h_t^2 &&
\end{align*}
Therefore,
\begin{align*}
P_{z,t} &= \sum_{i=1}^L P_{z,t}^i \leq \sum_{i=1}^L \frac{\delta}{(1+\delta)^{L-i+1} NP h_t^2} 
\end{align*}
or
\begin{align}
P_{z,t} & \leq NP h_t^2\left[1 - \frac{1}{(1+\delta)^L}\right]
\label{eqn:uppr_bnd_noise}
\end{align}

The results in \eqref{eqn:source_power_at_dstntn} and \eqref{eqn:noise_power_at_dstntn} imply that we have the following for the achievable rate at the destination
\begin{align*}
R_t &= \frac{1}{2} \log\left[1 + \frac{1}{(1+\delta)^L} \frac{N^2 P_L h_t^2}{P_{z,t} + \sigma^2} \right]
\end{align*}

Similarly, the source and noise power reaching the eavesdropper $e$ respectively are: 
\begin{align*}
P_{s,e} &= P_s h_s^2 \left(\prod_{i=1}^{L-1} N^2\beta_i^2 h_i^2\right) N^2 \beta_L^2 h_e^2= \frac{N^2 P_L h_e^2}{(1+\delta)^L}\\
P_{z,e} &= \sum_{i=1}^L P_{z,t}^i = \sum\limits_{i=1}^{L-1}N \beta_i^2 h_i^2 \left(\prod\limits_{j=i+1}^{L-1}(N \beta_j h_j)^2\right)N^2 \beta_L^2 h_e^2 + N \beta_L^2 h_e^2 = P_{z,t}\  h_e^2/h_t^2
\end{align*}
and the achievable rate at the eavesdropper is:
\begin{align*}
R_e &= \frac{1}{2} \log\left[1 + \frac{1}{(1+\delta)^L} \frac{N^2 P_L h_e^2}{P_{z,t} h_e^2/h_t^2 + \sigma^2} \right]
\end{align*}

Therefore, the achievable secrecy rate is
\begin{equation}
\label{eqn:hghSNRsecRate}
R_s = R_t - R_e = \frac{1}{2}\log\left[\frac{1 + \frac{1}{(1+\delta)^L} \frac{N^2 P_L h_t^2}{P_{z,t} + \sigma^2}}{1 + \frac{1}{(1+\delta)^L} \frac{N^2 P_L h_e^2}{P_{z,t} h_e^2/h_t^2 + \sigma^2}}\right]
\end{equation}

Thus, we have the following for the gap between the cutset upper-bound and the achievable secrecy rate
\begin{align}
C_{cut} - R_s &= \frac{1}{2}\log\left[\frac{1 + P_t/\sigma^2}{1 + P_e/\sigma^2}\right] - \frac{1}{2}\log\left[\frac{1 + \frac{1}{(1+\delta)^L} \frac{N^2 P_L h_t^2}{P_{z,t} + \sigma^2}}{1 + \frac{1}{(1+\delta)^L} \frac{N^2 P_L h_e^2}{P_{z,t} h_e^2/h_t^2 + \sigma^2}}\right]
\label{eqn:actual_gap}
\end{align}

Here, R.H.S. is an increasing function of $P_{z,t}$. Thus, from \eqref{eqn:uppr_bnd_noise} and \eqref{eqn:actual_gap}, we have

\begin{align}
C_{cut} - R_s &\leq \frac{1}{2}\log\left[\frac{\left(1+\left(1-\frac{1}{(1+\delta)^L}\right) \frac{NP h_t^2}{\sigma^2}\right) \left(1+\frac{N^2 P h_t^2}{\sigma^2}\right) \left(1+\left(1-\frac{1}{(1+\delta)^L}\right)\frac{N P h_e^2}{\sigma^2} + \frac{N^2 P h_e^2}{(1+\delta)^L \sigma^2}\right)}{\left(1+\left(1-\frac{1}{(1+\delta)^L}\right) \frac{NP h_e^2}{\sigma^2}\right) \left(1+\frac{N^2 P h_e^2}{\sigma^2}\right) \left(1+\left(1-\frac{1}{(1+\delta)^L}\right)\frac{N P h_t^2}{\sigma^2} + \frac{N^2 P h_t^2}{(1+\delta)^L \sigma^2}\right)}\right] \nonumber\\
			& \leq \frac{1}{2}\log\left[\!\frac{\left(1+L \delta\frac{N P h_t^2}{\sigma^2}\right)\left(1+\frac{N^2 P h_t^2}{\sigma^2}\right)\left(1+ L \delta\frac{N P h_e^2}{\sigma^2} + (1-L\delta) \frac{N^2 P h_e^2}{\sigma^2}\right)}{\left(1+L \delta\frac{N P h_e^2}{\sigma^2}\right)\left(1+\frac{N^2 P h_e^2}{\sigma^2}\right)\left(1+ L \delta\frac{N P h_t^2}{\sigma^2} + (1-L\delta) \frac{N^2 P h_t^2}{\sigma^2}\right)}\right] \quad\left[\mbox{as }{1}/{(1+\delta)^L} \ge 1-L \delta\right]\nonumber\\
			&= \frac{1}{2}\log\left[\left.\frac{1+{N^2 P h_t^2}/{\sigma^2}}{1+\frac{(1-L\delta) NPh_t^2/\sigma^2}{1+L\delta P h_t^2/\sigma^2}}\middle/ \right. \frac{1+{N^2 P h_e^2}/{\sigma^2}}{1+\frac{(1-L\delta) NPh_e^2/\sigma^2}{1+L\delta P h_e^2/\sigma^2}}\right]\nonumber\\
			&\leq \frac{1}{2}\log\left[\left(\left. \frac{1+L\delta NP h_t^2/\sigma^2}{(1-L\delta)}\right)\middle/ \right.\left(1+L\delta \frac{NPh_e^2}{\sigma^2}\right)\right]\nonumber\\
			&=\frac{1}{2}\log\left[\frac{1}{(1-L\delta)} \frac{1+L\delta NP h_t^2/\sigma^2}{1+L\delta NP h_e^2/\sigma^2}\right] \label{eqn:gap_bnd}
\end{align}

Since, $R_{s,opt} \geq R_s, \  C_{cut}-R_{s,opt} \leq C_{cut}-R_{s}$. Note that as $\delta \rightarrow 0, \ C_{cut}-R_s \rightarrow 0$, i.e., secrecy rate approaches the cut-set bound. 

\section{Numerical Simulations}
\label{sec:numSim}
\begin{figure}[!t]
\begin{subfigure}[b]{0.5\textwidth}
\includegraphics[width=\textwidth]{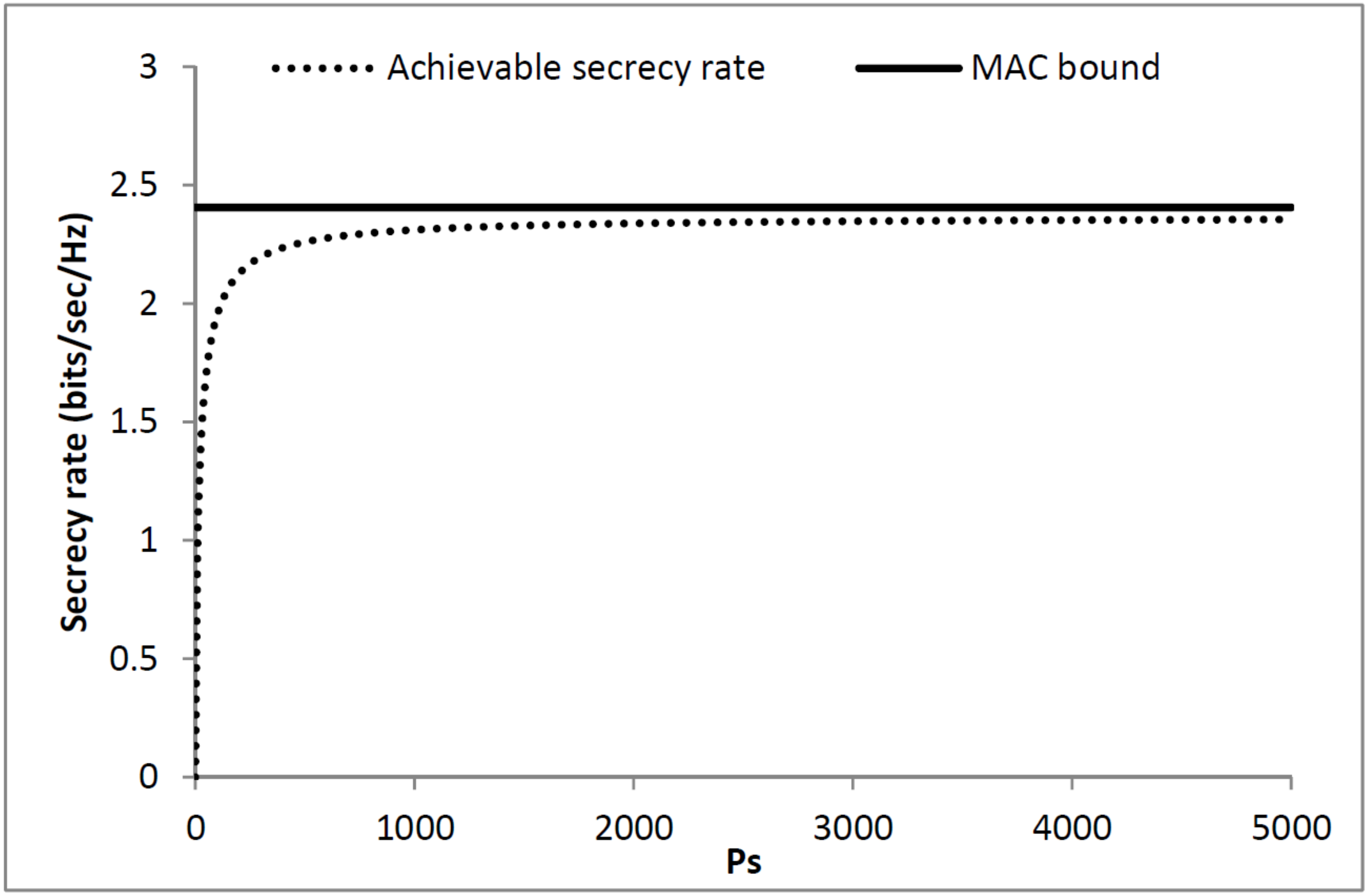}
\caption{}
\label{fig:2lyrdNetSecRateHighSNRplot1}
\end{subfigure}
\begin{subfigure}[b]{0.5\textwidth}
\includegraphics[width=\textwidth]{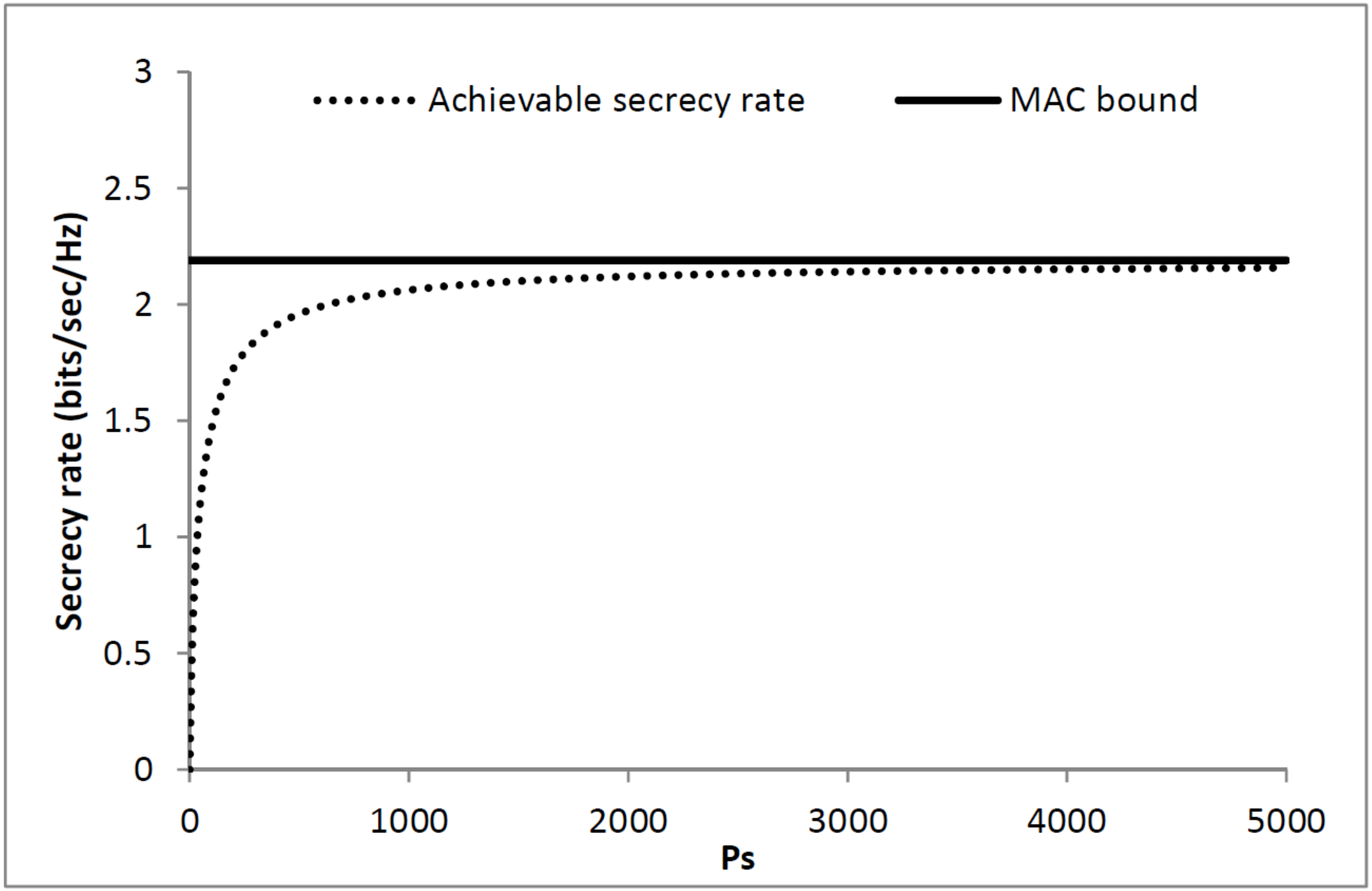}
\caption{}
\label{fig:2lyrdNetSecRateHighSNRplot2}
\end{subfigure}
\caption{Plot of achievable secrecy rate with all relays transmitting at maximum power against varying source power along with the cutset bound  for $2$-layer network with $2$ nodes in each layer.  $P=500,\   \sigma^2=1.0, \ \delta=0.005,\ $ (a)~$ h_s=0.689, \  h_1=0.603, \  h_t=0.203, \   h_e= 0.031,$  and (b)~$ h_s=0.260,  \  h_1 = 0.925,\   h_t=0.113, \    h_e=0.012.$}
\label{fig:2lyrdNetSecRateHighSNRplot}
\end{figure}

In this section, we present numerical results to evaluate the performance of the proposed high SNR approximation scheme. We consider a 2-layer network with two nodes in each layer and eavesdropper snooping on the transmissions of the nodes in the last layer. In Figure~\ref{fig:2lyrdNetSecRateHighSNRplot}, we plot the achievable secrecy rate when all the nodes transmit at their maximum power along with the corresponding cut-set as a function of source power for the specified system parameters.

From the figure, it can be observed that at high SNR the achievable rate lies within a constant gap from the cut-set bound. In Figure~\ref{fig:2lyrdNetSecRateHighSNRplot1}, the actual gap between the cut-set bound and achievable rate is $0.05$ bits/sec/Hz while the upper bound on this gap as given by \eqref{eqn:gap_bnd} is $0.25$ bits/sec/Hz. Similarly, in Figure~\ref{fig:2lyrdNetSecRateHighSNRplot2}, the actual gap is $0.03$ bits/sec/Hz while the upper bound on this gap is $0.09$ bits/sec/Hz. Thus, it can be seen that \eqref{eqn:gap_bnd} tightly approximates the gap between achievable secrecy rate and the cut-set bound. 

\section{Conclusion and Future Work}
\label{sec:cnclsn}
We consider the problem of secure ANC rate maximization over a class of Gaussian layered networks where a source communicates with a destination through $L$ intermediate relay layers with $N$ nodes in each layer in the presence of a single eavesdropper which can overhear the transmissions of the nodes in any one layer. The key contribution of is the computation of the globally optimal set of scaling factors for the nodes that maximizes the end-to-end secrecy rate for a class of layered networks. We also show that in the high-SNR regime, ANC achieves secrecy rates within a constant gap of the cutset upper bound on the secrecy capacity and numerically validate this. In future, we plan to extend this work for more general layered networks.

\appendices

\section{Proof of Lemma~\ref{lemma:diamondNetReducedBeta1}}
\label{appndx:lemma0}
\setcounter{equation}{0}
\renewcommand{\theequation}{\thesection.\arabic{equation}}
To find the value of $\bm{\beta}_{opt}$ that maximizes the corresponding secrecy rate, equating the partial derivative of the secrecy rate in \eqref{eqn:secRate4lemma2} with respect to $\beta_{i}$ to zero, we get a system of $N$ simultaneous polynomial equations. Without any loss of generality, subtracting the equation corresponding to the partial derivative with respect to $\beta_{1}$ from the rest of $N-1$ equations, it is easy to prove that $\beta_{1} = \beta_{2} = \cdots = \beta_{N} = \beta$ is the only root of this system of equations. Substituting this solution in one of the equations, we get
\begin{equation*}
\beta \left(N^2 h_t^2 h_e^2 (N P_s h_s^2 + \sigma^2)\beta^4 - \sigma^2\right) = 0
\end{equation*}
This equation has the following two distinct and real solutions for the stationary points of the secrecy rate with respect to $\beta$:
\begin{equation*}
\beta_{z} = 0, \mbox{ and } \beta_{glb}^2 = \sqrt{\frac{1}{N^2 h_t^2 h_e^2 \left(1+N\frac{P_s h_s^2}{\sigma^2} \right)}}
\end{equation*}

Now using the second derivative test, we can prove that when $h_t>h_e$, then $\beta_{z}$ and $\beta_{glb}$ are the points of global minimum and maximum, respectively. Similarly, for $h_e > h_t$, $\beta_{z}$ and $\beta_{glb}$ can be proved to be the points of global maximum and minimum, respectively. Given the convex nature of the secrecy rate function with respect to $\beta$ for $h_t>h_e$ and that $\beta_{max}$ is the largest value of the scaling factor $\beta$, we have the result
\begin{equation*}
\beta_{opt}^2 = \min(\beta_{max}^2, \beta_{glb}^2), \quad \mbox{if } h_t > h_e
\end{equation*}

\section{Proof of Lemma~\ref{lemma:lyrdNetReducedBeta1}}
\label{appndx:lemma4}
\setcounter{equation}{0}
\renewcommand{\theequation}{\thesection.\arabic{equation}}
From \eqref{eqn:eqProb}, we have
\begin{align*}
\bm{\beta}_{opt} &= \argmax_{\bm{\beta}} \frac{1+SNR_t}{1+SNR_e} \\
                 &= \argmax_{(\bm{\beta}_{1}, \ldots, \bm{\beta}_{M})} \frac{\argmax\limits_{(\bm{\beta}_{M+1}, \ldots, \bm{\beta}_{L})} 1+SNR_t}{1+SNR_e} 
\end{align*}
where the last step follows from $SNR_e$ being only a function of $(\beta_{1}, \ldots, \beta_{M})$.

Using \cite[Lemma 2]{112agnihotriJaggiChen}, it is straightforward to establish that
\begin{equation*}
(\bm{\beta}_{M+1,max}, \ldots, \bm{\beta}_{L,max}) = \argmax\limits_{(\bm{\beta}_{M+1}, \ldots, \bm{\beta}_{L})} 1+SNR_t,
\end{equation*}
where we have
\begin{align*}
\beta_{M+1,n,max}^2 &= \frac{P_{M+1,n}}{P_{Rx,M+1}}, n \in \{1, \ldots, N\}\\
\beta_{i,n,max}^2 &= \frac{P_{i,n}}{P_{Rx,i}}, \quad i \in \{M+2, \ldots, L\}, n \in \{1, \ldots, N\},
\end{align*}
where
\begin{align*}
P_{Rx,M+1} &= h_s^2 H_{1,M-1}^2 P_s \left(\sum_{n=1}^N \beta_{M,n}\right)^2 h_M^2+\left[\left(\sum_{i=1}^{M-1} G_{i,M-1}^2\right)\left(\sum_{n=1}^N \beta_{M,n}\right)^2 h_M^2+\left(\sum_{n=1}^N \beta_{M,n}^2\right) h_M^2\right] \sigma^2+\sigma^2\\
           &= S_{M+1}P_s+(N_{1:M}+1)\sigma^2\\
P_{Rx,i}   &= S_{M+1}P_s H_{M+1,i}^2+N_{1:M+1} H_{M+1,i}^2+ \sigma^2\left[\sum_{j=M+1}^{i-1} G_{j,i-1}^2+1\right]
\end{align*}

Note that as in the ECGAL networks, all channels gains between two adjacent layers are equal, the received power at all relays in a layer is the same.

\section{Proof of Lemma~\ref{lemma:lyrdNetReducedBeta2}}
\label{appndx:lemma5}
\setcounter{equation}{0}
\renewcommand{\theequation}{\thesection.\arabic{equation}}
To find the value of $\bm{\beta}_{M,opt}$ that maximizes the corresponding secrecy rate, equating the partial derivative of the secrecy rate in \eqref{eqn:secRate4lemma2} with respect to $\beta_{M,n}$ to zero, we get a system of $N$ simultaneous polynomial equations. Without any loss of generality, subtracting the equation corresponding to the partial derivative with respect to $\beta_{M,1}$ from the rest of $N-1$ equations, it is easy to prove that $\beta_{M,1} = \beta_{M,2} = \cdots = \beta_{M,N} = \beta_{M}$ is one of the roots of this system of equations. Substituting this solution in one of the equations, we get
\begin{equation*}
\beta_M \left({\mathcal A}\,\beta_M^{4}-|{\mathcal B}|\,\beta_M^{2}+{\mathcal C}\right) = 0
\end{equation*}
This equation has the following three distinct and real solutions for the stationary points of the secrecy rate with respect to $\beta_M$: $\beta_{M,z} = 0$ and
\begin{equation*}
\beta_M^2 = \begin{cases}
              \frac{|{\mathcal B}|}{2\,|{\mathcal A}|}\left(\sqrt{1 + \frac{4\,|{\mathcal A}|\,{\mathcal C}}{{\mathcal B}^2}} - 1\right), {\mathcal A} < 0, {\mathcal C} > 0\\
             \frac{|{\mathcal B}|}{2\,{\mathcal A}}\left(\sqrt{1 + \frac{4\,{\mathcal A}\,|{\mathcal C}|}{{\mathcal B}^2}} + 1\right), {\mathcal A} > 0, {\mathcal C} < 0
          \end{cases}
\end{equation*}

Before discussing the nature of these stationary points, note that
\begin{footnotesize}
\begin{align*}
h_M^{2} \alpha\!-\!h_e^{2} \nu \!> \!0 \!\implies \!h_e^{2} \alpha \nu (2 F\!+\!1) \left(\!2 F\!+\!1\!+\!2 \frac{P_s}{\sigma^2} E\right)\!-\!h_M^{2} [2 \lambda E \!+\!(2 F\!+\!1)\mu] \left[\!(2 F\!+\!1)\mu\!+\!2 \left(\!\lambda\! +\!\alpha \frac{P_s}{\sigma^2}\right) E \right]\! <\! 0\\
h_e^{2} \alpha \nu (2 F\!+\!1) \left(\!2 F\!+\!1\!+\!2 \frac{P_s}{\sigma^2} E\right)\!-\!h_M^{2} [2 \lambda E \!+\!(2 F\!+\!1)\mu] \left[\!(2 F\!+\!1)\mu\!+\!2 \left(\!\lambda \!+\!\alpha \frac{P_s}{\sigma^2}\right)\! E \right]\! >\! 0 \! \implies \!h_M^{2} \alpha\!-\!h_e^{2} \nu\! <\! 0
\end{align*}
\end{footnotesize}

Now using the second derivative test, we can prove that when $h_M^{2}\,\alpha-h_e^{2}\,\nu > 0$, then $\beta_{M,z}$ and $\beta_{M,1}$ are the points of global minimum and maximum, respectively. Similarly, for $h_e^{2} \alpha \nu (2 F+1) \left(2 F+1+2 \frac{P_s}{\sigma^2} E\right)-h_M^{2} [2 \lambda E +(2 F+1)\mu] \left[(2 F+1)\mu+2 \left(\lambda +\alpha \frac{P_s}{\sigma^2}\right) E \right] > 0$, $\beta_{M,z}$ and $\beta_{M,2}$ can be proved to be the points of global maximum and minimum, respectively. To emphasize that $\beta_{M,1}$ is the point of global maximum of the secrecy rate, we rechristen it $\beta_{M,glb}$. Given the convex nature of the secrecy rate function with respect to $\beta_M$ for $h_M^{2}\,\alpha-h_e^{2}\,\nu > 0$ and that $\beta_{M,max}$ is the largest value of the scaling factor $\beta_M$, we have the result
\begin{equation*}
\beta_{M,opt}^2 = \min(\beta_{M,max}^2, \beta_{M,glb}^2), \quad \mbox{if } h_M^{2}\,\alpha-h_e^{2}\,\nu > 0
\end{equation*}

\section{Proof of Lemma~\ref{lemma:lyrdNetReducedBeta3}}
\label{appndx:lemma6}
With optimum values of the scaling factors for the nodes in layers $(M, M+1, \ldots, L)$ from Lemmas \ref{lemma:lyrdNetReducedBeta1} and \ref{lemma:lyrdNetReducedBeta2}, the problem \eqref{eqn:eqProb} of computing the optimal network-wide scaling vector reduces to
\begin{align*}
\bm{\beta}_{opt} &= \argmax_{(\beta_{1}, \ldots, \beta_{M-1}, \beta_{M,opt}, \ldots, \beta_{N,max})} \frac{1+SNR_t}{1+SNR_e}
\end{align*}
Similar to \cite[lemma 3]{116agrawalAgnihotri} for linear chain networks, we can show that $\frac{1+SNR_t}{1+SNR_e}$ is a quasi-convex function of $\bm{\beta}_{M-1}$ in the interval $[-\bm{\beta}_{M-1,max}, \bm{\beta}_{M-1,max}]$ for a given sub-vector $(\bm{\beta}_1, \ldots, \bm{\beta}_{M-2})$ of scaling factors of first $M-2$ relay layers and optimum sub-vector $(\beta_{M,opt}, \ldots, \beta_{L,opt})=(\beta_{M,opt}, \ldots, \beta_{L,max})$ of the optimum scaling factors of the remaining relays. Thus, $\bm{\beta}_{M-1,opt} = \bm{\beta}_{M-1,max}$. Carrying out this process successively for relays in layer $M-2, \ldots, 1$, proves the lemma.

\end{document}